\documentclass[conference]{IEEEtran}

\IEEEoverridecommandlockouts

\usepackage[noadjust]{cite}
\usepackage{amsmath,amssymb,amsfonts,amsthm}
\usepackage{algorithmic}
\usepackage{enumitem}
\usepackage{graphicx}
\usepackage{subfigure}
\usepackage{textcomp}
\usepackage{multirow}
\usepackage{listings}
\usepackage{pgfplots}
\usepackage{float}
\pgfplotsset{compat=1.18}
\usepackage[table]{colortbl}
\usepackage[misc,geometry]{ifsym}
\usepackage{hyperref}
\usepackage{xspace}

\newcommand{\para}[1]{\smallskip \noindent{\bf #1}}

\theoremstyle{definition}
\newtheorem*{example}{Example}

\definecolor{colorGk}{RGB}{55,126,184}  % 蓝色系 for gen_keys
\definecolor{colorP}{RGB}{255,127,0}   % 橙色系 for proving
\definecolor{colorV}{RGB}{77,175,74}   % 绿色系 for verifying
\definecolor{plum}{RGB}{148, 0, 211}

\newcommand{\ie}{{\em i.e.,}\xspace}

\newcommand{\eg}{{\em e.g.,}\xspace}

\newcommand{\fig}{{Fig.}}

\begin{document}

\title{Zero-Knowledge Verifiable Graph Query Evaluation via Expansion-Centric Operator Decomposition}

\author{
\IEEEauthorblockN{Hao Wu$^1$, Changzheng Wei$^{1,2}$, Yanhao Wang$^1$, Li Lin$^{2}$, Yilong Leng$^1$, \\ Shiyu He$^1$, Minghao Zhao$^{1*}$\thanks{$^*$Corresponding authors.},
Hanghang Wu$^2$, Ying Yan$^{2*}$, and Aoying Zhou$^1$}
\IEEEauthorblockA{$^1${\em School of Data Science and Engineering, East China Normal University} \quad $^2${\em Digital Technologies, Ant Group}\\
51265903006@stu.ecnu.edu.cn,\; changzheng.wcz@antgroup.com,\;  \\ yhwang@dase.ecnu.edu.cn,\; felix.ll@antgroup.com
\{51265903087, 51255903085\}@stu.ecnu.edu.cn,\; \\ mhzhao@dase.ecnu.edu.cn,\; \{hanghang.whh, fuying.yy\}@antgroup.com,\; ayzhou@dase.ecnu.edu.cn}
}

\maketitle

\begin{abstract}
This paper investigates the feasibility of achieving zero-knowledge verifiability for graph databases, enabling database owners to cryptographically prove the query execution correctness without disclosing the underlying data.
Although similar capabilities have been explored for relational databases, their implementation for graph databases presents unique challenges.
This is mainly attributed to the relatively large complexity of queries in graph databases.
When translating graph queries into arithmetic circuits, the circuit scale can be too large to be practically evaluated.

To address this issue, we propose to break down graph queries into more fine-grained, primitive operators, enabling a step-by-step evaluation through smaller-scale circuits.
Accordingly, the verification with ZKP circuits of complex graph queries can be decomposed into a series of composable cryptographic primitives, each designed to verify a fundamental structural property such as path ordering or edge directionality.
Especially, having noticed that the graph expansion (\ie traversing from nodes to their neighbors along edges) operation serves as the backbone of graph query evaluation, we design the {\em expansion-centric operator decomposition}.
In addition to constructing circuits for the expansion primitives, we also design specialized ZKP circuits for the various attributes that augment this traversal.
The circuits are meticulously designed to take advantage of PLONKish arithmetization.
By integrating these optimized circuits, we implement ZKGraph, a system that provides verifiable query processing while preserving data privacy.
Performance evaluation indicates that ZKGraph significantly outperforms naive in-circuit implementations of graph operators, achieving substantial improvements in both runtime and memory consumption.
\end{abstract}


\section{Introduction}

Graph databases (\eg Neo4j, Amazon Neptune, and TuGraph) model data as graph structures, where nodes represent entities and edges define their relationships.
This architecture provides significant advantages for querying and analyzing complex interconnected datasets.
Owing to their ability to efficiently traverse relationships, graph databases have gained widespread adoption in various domains, including social network analysis\cite{10.1145/2484425.2484442}, financial transaction processing\cite{9424591}, and bioinformatics network analysis\cite{10.1093/bioinformatics/btx397}.

As graph databases gain growing adoption in data-sensitive domains, a critical need arises to ensure both query correctness and data privacy simultaneously.
In such a scenario, the {\em data owner} maintains the dataset stored in a graph database, and a {\em data user} submits queries (\eg using Cypher statements) to retrieve specific results.
The data user needs to be convinced that the query result he received is the result of executing the query correctly on the dataset the data owner declared, and the data owner must prove it without revealing the dataset to the data user.
More formally, this means providing the {\em zero-knowledge verifiability} on graph query evaluation, in which the data user can {\em cryptographically verify} that the query results are correctly computed by executing the specified operations on the data owner's declared dataset, whereas the data owner provides such a proof without disclosing dataset information.

%A data user submits queries (e.g., via Cypher statements) to retrieve specific results
%Such datasets often contain highly sensitive information, 
%presenting a fundamental challenge for data providers: they must protect data privacy while simultaneously providing users with strong assurances about the integrity of query results.
%Consequently, a critical need has emerged for systems that can execute queries over private graph datasets and, in turn, provide verifiable proofs of the results' correctness and completeness.
%In this setting, a client must be able to trust the results returned by a server that holds the private graph, without needing direct access to the graph itself.

For example, consider an analyst on a social network platform tasked with identifying key influencers in a viral cascade or verifying the community structure around a sensitive topic.
Accomplishing these tasks requires the execution of complex graph queries (\eg reachability and shortest path) on the raw, sensitive graph data.
Using the platform's conventional, opaque analytics systems, it is difficult for the analyst to independently verify the correctness and completeness of the query results.
Conversely, if the platform grants the analyst direct query access to the raw, sensitive graph data, it risks unacceptable disclosure of user information.

%%%%Zero-knowledge proofs (ZKPs) \cite{goldreich1991proofs,goldwasser1985knowledge} provide a powerful cryptographic paradigm for addressing the above challenge.
%%%%%%%Specifically, they enable a prover (\eg a graph database owner) to convince a verifier (\eg a client) of the validity of a computation result without revealing any information beyond the result itself.
%%%%%%%%%%%%%ZKPs can be categorized into interactive ZKPs and non-interactive ZKPs %(NIZKPs).
%%%%%%The requisite multi-round interaction in interactive ZKP imposes burdensome availability constraints and yields non-transferable proofs, precluding asynchronous or public verification.
%In contrast, NIZKPs allow the prover to generate a single, self-contained proof that can be verified by anyone possessing the verification key.
%This support for public verifiability and offline verification is particularly advantageous for building verifiable database systems.
%Consequently, a significant body of research has been dedicated to developing and optimizing NIZKPs \cite{blum1988non,groth2016size,maller2019sonic,cryptoeprint:2019/1021}.

In cryptography, Zero-Knowledge Proofs (ZKP) \cite{goldreich1991proofs,goldwasser1985knowledge} offer a powerful tool to realize such functionality.
It is a cryptographic protocol that enables a prover to convince a verifier that a given statement is true, without revealing any information beyond the validity of the statement itself.
%They enable a prover (\eg a graph database owner) to convince a verifier (\eg a client) of the validity of a computation's result without revealing any information beyond the result itself. 
In recent years, researchers have explored the integration of zero-knowledge proofs (ZKPs) into database systems to simultaneously achieve verifiable query processing and data privacy preservation \cite{li2023zksql,10.1145/3709713}.
In these works, the database operators, including relational algebra operators such as selection ($\sigma$) and derived operators like join ($\Join$), are first translated into ZK circuits, the foundational computational units in ZKP systems, and then these circuits are cryptographically evaluated to generate privacy-preserving query proofs.

% Firstly, it simplifies the verification process, allowing clients to check proofs offline at their convenience.
% Secondly, NIZK inherently supports public verifiability, where a single proof can be published and verified by multiple independent parties, fostering broader trust.
% Recent years have witnessed significant advancements in both the theory and practice of NIZK, laying the groundwork for building verifiable computation systems.
% Consequently, our work, ZKGraph, focuses on leveraging NIZK to achieve verifiable graph query evaluation.

However, this approach cannot be directly applied to graph databases due to fundamental differences in computational paradigms. Unlike relational operators, graph operators are inherently algorithmic (\eg neighborhood expansion, reachability, and pathfinding), requiring iterative computation (multi-hop traversals), dynamic data access (data-dependent branching), or irregular data structures (unpredictable connectivity).
{\em These properties conflict with the static circuit model of zero-knowledge proofs}, where computations must be {\em fixed-structure} (predefined execution paths) and {\em data-independent} (deterministic memory access).
Consequently, the native implementations yield prohibitive proving overhead.

To address this challenge, we introduce the {\em expansion-centric} operator decomposition method.
It breaks down graph queries into more fine-grained, primitive operators, enabling a step-by-step evaluation through smaller, more manageable ZK circuits.
Accordingly, the verification of a complex graph query is transformed into a chain of composable cryptographic primitives.
Consequently, the verification burden of proving a dynamic and complex computational process is shifted to validating a sequence of well-defined, static properties about the query’s intermediate and final results.
The cornerstone of this model is the expansion operation, the fundamental action of navigating from a node to its neighbors.
Most graph operations, such as finding the shortest path, filtering by properties, or ordering results, can be conceptualized as compositions or attribute-laden variations of this core expansion primitive.
This expansion-centric decomposition aligns well with the ZKP arithmetization. 
Instead of executing the entire graph query within a circuit, our approach uses the circuit to efficiently verify that the intermediate and final results are consistent with both the query's constraints and the private graph database.

Based on the expansion-centric model, we present ZKGraph\footnote{The source code, data, and/or other artifacts have been made available at \url{https://github.com/Hao8172/ZKGraph}}, which, to the best of our knowledge, is the first system to leverage non-interactive zero-knowledge proofs for the confidential and verifiable evaluation of arbitrary graph queries.
ZKGraph materializes our expansion-centric philosophy by translating fundamental graph operations into a suite of highly optimized, composable PLONKish circuits \cite{zcash_plonkish}, such as node neighborhood expansion and single-source shortest path computation.
By composing these foundational circuits, ZKGraph can support various complex graph queries, ensuring both the privacy of the data and the verifiability of the results.

Finally, we evaluate the performance of ZKGraph on the LDBC SNB benchmark.
The results demonstrate that ZKGraph significantly outperforms naive in-circuit implementations of graph operators, achieving substantial improvements in both runtime and memory consumption.

% Graph databases have become indispensable for storing and analyzing complex interconnected data, such as social networks, financial transactions, and bioinformatics networks.
% In this paper, we propose ZKGraph, a system that leverages non-interactive zero-knowledge proofs to support the confidential and verifiable querying of graph data. 
% To the best of our knowledge, it is the first system specifically designed to offer Zero-Knowledge verifiable computation tailored for graph queries.
% The primary contribution of ZKGraph lies in the design of specially optimized ZKP circuits for fundamental graph query operations, such as node neighborhood expansion and single-source shortest path computations. 
% These circuits, custom-tailored for the nuances of graph computations, not only consider the correctness of the computation but also emphasize efficiency and succinctness within the zero-knowledge context. 
% By meticulously designing these foundational graph primitive circuits, ZKGraph can flexibly combine them to support broader and more complex graph queries, all while ensuring the privacy of the querying process and the verifiability of the results. 

\section{Background}

\subsection{Zero-Knowledge Proof}

A zero-knowledge proof (ZKP) is a cryptographic protocol that enables one party (\ie the prover) to convince another party (\ie the verifier) that a given statement is true without conveying any additional information beyond the validity of the statement itself.
For example, let us consider a scenario in which a data owner possesses a private graph database (\ie the witness) and wishes to assure an analyst that a specific path exists between two nodes.
Using ZKP, the owner can prove the existence of this path without disclosing any details about the graph's structure, other nodes, and alternative paths.

A fundamental distinction among ZKP protocols lies in the requirement for interaction between the prover and the verifier.
Interactive ZKP systems require multiple rounds of communication to establish the proof's validity.
This interactive nature can be cumbersome and impractical in some real-world scenarios, particularly in multi-verifier settings, which are inherently impractical as the prover must conduct a separate, full session with each party.
% particularly in the client-server architecture, where the verifier may not always be online.
% Moreover, a proof generated in an interactive session is valid only for the specific verifier who participates.
% Consequently, a separate, full interactive session is required for each new verifier who wishes to validate the statement.

In contrast, non-interactive ZKP (NIZKP) can address these limitations.
Using techniques such as the Fiat-Shamir heuristic \cite{10.1007/3-540-47721-7_12} and the common reference string \cite{blum1988non}, a prover can generate a single self-contained cryptographic proof within only one round of communication.
This single proof can be published or broadcast, allowing any number of verifiers possessing the verification key to independently check its validity at any time.
Hence, NIZKPs are well-suited for applications such as verifiable databases, where the server must provide the client with a verifiable result that can be checked offline.
This paper will focus solely on NIZKPs for verifiable graph databases.

\subsection{PLONKish Arithmetization}

NIZKPs typically express the computation to be proved as a system of polynomial constraints, known as the \emph{arithmetization} process.
Our methodology adopts the PLONKish arithmetization \cite{zcash_plonkish} implemented in Halo2, which represents the computation as a two-dimensional table of values.
The table is composed of several types of columns:
\begin{itemize}
    \item \textbf{Advice Columns} hold private witness values provided by the prover, such as private input and intermediate values computed during execution.
    \item \textbf{Fixed Columns} hold constant values to define the circuit's logic, such as selectors to activate specific rules on certain rows and predefined constants for the computation. They are known to the prover and the verifier.
    \item \textbf{Instance Columns} hold public input and output of the computation, which are also known to both parties.
\end{itemize}
The relationship between the values within cells is enforced by a set of rules called \emph{gates} or \emph{constraints}. 
These constraints are polynomial equations that must hold for each row of the circuit.
The selectors in fixed columns are used to designate which gate applies to which row, allowing a single circuit to support various operations.
By enforcing the constraints, the ZKP system ensures the computational integrity of the prover's statement without revealing the underlying witness.

\begin{figure}[t]
    \centering
    \includegraphics[width=0.8\linewidth]{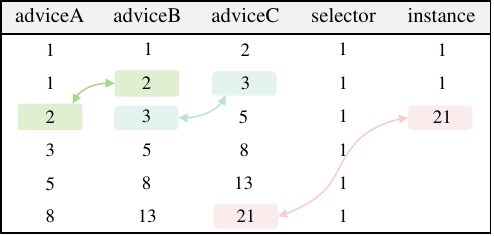}
    \caption{Illustration of a PLONKish Fibonacci circuit.}
    \label{fig:PLONKish_Fibonacci_Circuit}
\end{figure}

\begin{example}
\fig~\ref{fig:PLONKish_Fibonacci_Circuit} illustrates a PLONKish circuit for the classic Fibonacci problem.
Specifically, it proves that for a Fibonacci sequence beginning with $f(0) = 1$ and $f(1) = 1$, the 8th term, $f(8)$, is indeed $21$.
Three advice columns, namely \emph{adviceA}, \emph{adviceB}, and \emph{adviceC}, are used to store the intermediate values of the computation.
The instance column holds the public data that the verifier can access, which includes the public input $(1, 1)$ and the prover's claimed result of $21$.
To enforce the Fibonacci rule, a selector activates an addition gate for each row.
This gate is a polynomial constraint that must be satisfied for the proof to be valid, \ie $S[i] * (A[i] + B[i] - C[i]) = 0$.
This ensures that, for every step, the sum of the first two advice columns equals the third.

Moreover, the circuit must ensure the correct flow of data between different rows, which is achieved through equality checks.
As visualized by the arrows in the diagram, these constraints enforce that the output of one row becomes the input of the next.
They create the ``wiring'' of the circuit, forcing the value from adviceB in one row to be equal to the value from adviceA in the subsequent row, and likewise for adviceC and adviceB.
Such a propagation guarantees that the sequence is built correctly from the initial state to the final term, which is itself linked via an equality check to the claimed value in the public instance column.
\end{example}

\subsection{Graph Database Operators}
Graph query processing relies on a suite of fundamental operators to navigate and analyze complex interconnected data. 
These operators provide rich semantics, including traversal for navigating from nodes to their neighbors, filtering to selectively include or exclude entities based on their properties, path-finding to identify routes between nodes (such as the shortest path), and aggregation to compute summary statistics like counting degrees or averaging property values.

In the context of verifiable computation, the primary challenge lies in creating efficient and composable cryptographic counterparts for these operators.
Their inherent complexity, characterized by iterative processes and data-dependent control flows over irregular data structures, fundamentally clashes with the static nature of ZKP arithmetization.
A direct translation of these algorithms into circuits would lead to prohibitively high proof generation costs.
Our proposed ``expansion-centric'' model is designed to address this specific challenge by deconstructing complex queries into a series of verifiable, modular steps.

\section{System Overview}
\label{sec:system}

This section outlines the architecture of ZKGraph.
We begin by defining the system goal and the security model, along with the cryptographic guarantees that our system provides.
Next, we present the workflow by which the system processes graph queries to provide the client with an authenticated answer.
To address the fundamental challenges of applying ZKPs to graph computations, we then introduce the ``expansion-centric'' model.
Finally, we discuss the rationale for our design choices regarding the edge storage format.

\subsection{System Goal}

The primary objective of ZKGraph is to enable verifiable query evaluation over private graph databases.
This process involves the following two entities:
\begin{itemize}
    \item \textbf{Prover:} The party that owns the private graph database and is responsible for executing queries and generating ZKPs of their correct computation.
    \item \textbf{Verifier:} The party that issues queries and verifies the accompanying proofs to ensure the integrity and correctness of the results received from the prover.
\end{itemize}
The system is designed to provide the verifier with cryptographic guarantees regarding the correctness and completeness of a query result, without compromising the confidentiality of the prover's underlying data.

\subsection{Security Model}

ZKGraph leverages the Halo2 proving system \cite{halo2}, a framework for constructing NIZKPs based on PLONKish arithmetization.
The system ensures the following properties:
\begin{itemize}
    \item \textbf{Correctness:} If an honest prover holds the authentic graph database and correctly generates the circuit of a query, they can always construct a proof that convinces an honest verifier of the statement's validity.
    \item \textbf{Soundness:} For any statement regarding a graph query that is not correct, the probability that a cheating prover can convince an honest verifier is negligible.
    \item \textbf{Knowledge Soundness:} When an honest verifier is convinced that a graph query statement is correct, it implies that the prover possesses a valid witness for the execution of the query on the committed graph that yields the claimed result.
    \item \textbf{Zero-Knowledge:} The verifier learns nothing beyond what can be inferred from the correctness of the specific graph query result and the fact that the prover knows a valid witness for its computation against the committed graph database.
\end{itemize}

Building upon the PLONKish arithmetization, ZKGraph translates graph queries into custom circuits.
The prover then assigns the witness to these PLONKish circuits.
This witness comprises the private graph data for the query and all intermediate values generated during its execution.
These circuits, which have been instantiated with the specific witness, serve as input to the Halo2 proving system.

Subsequently, these circuits are processed through the following key stages to generate a proof of the correct query execution.
Initially, the prover encodes the witness data into a set of polynomials and commits to them using polynomial commitments.
To demonstrate that all circuit constraints are satisfied, the prover constructs a vanishing argument, which consolidates all gate constraints into a single polynomial that must evaluate to zero across its domain if the computation is valid.
Following this, the prover computes the evaluations of various circuit polynomials at random challenge points derived non-interactively via the Fiat-Shamir heuristic.
To efficiently verify that these evaluations are consistent with their respective commitments, a multi-point opening argument is constructed, aggregating multiple opening claims into one.
Finally, the prover generates a succinct opening proof for this aggregate argument polynomial using the Kate-Zaverucha-Goldberg (KZG) commitment scheme \cite{10.1007/978-3-642-17373-8_11}.

We choose the KZG polynomial commitment scheme for its notable succinctness.
A key advantage of KZG is its provision of constant-size proofs and constant-time verification.
This property ensures that the proof length and verifier workload remain independent of the complexity of circuits, a critical feature for the scalability of our zero-knowledge graph databases.

\para{Cryptographic Guarantees.}
The correctness and security of ZKGraph are based on the well-established cryptographic properties inherent to Halo2 and our design of graph query-specific PLONKish circuits, as will be shown in Section~\ref{sec:oper}.

The security of the Halo2 protocol \cite{halo2book} is established under the assumption that all participants are bounded by probabilistic polynomial-time algorithms and fundamentally relies on the hardness of the discrete log relation problem within its specified prime-order cyclic group.
Halo2 is designed to provide robust zero-knowledge arguments by satisfying several critical security notions.

Halo2 achieves perfect completeness, ensuring that an honest prover possessing a valid witness can always convince the verifier.
Soundness, which prevents a cheating prover from validating an incorrect statement, is analyzed through the lens of state-restoration soundness.
This is particularly pertinent as Halo2 protocols are typically rendered non-interactive using the Fiat-Shamir transformation, with security considered in the random oracle model.
Knowledge soundness is demonstrated via witness-extended emulation, notably within the Algebraic Group Model (AGM) \cite{10.1007/978-3-319-96881-0_2}, which facilitates the extraction of a witness from any successful algebraic state-restoration prover, often without requiring prover rewinding.
Furthermore, Halo2 provides perfect special honest-verifier zero-knowledge, which is substantiated by the existence of a simulator capable of producing interaction transcripts that are computationally indistinguishable from those of a real prover-verifier exchange for any polynomial-time decidable relation.

Moreover, ZKGraph's circuit implementations are specifically designed for oblivious execution.
This means that the sequence of operations performed by the prover is independent of the actual values within the private graph database.
The characteristic of data-independent processing is crucial for preventing any leakage of information about the private database beyond the query result itself.
For example, the \textit{expansion} operator is designed to execute a fixed pattern of computations, regardless of whether specific graph elements satisfy the query conditions.
To address the challenge of protecting the privacy of data sizes, ZKGraph adopts the method of ``dummy tag'' introduced in ZKSQL \cite{li2023zksql}, which involves the insertion of dummy elements into its query evaluation process to effectively obscure the actual data sizes.

\subsection{System Workflow}

\begin{figure}[t]
    \centering
    \includegraphics[width=\linewidth]{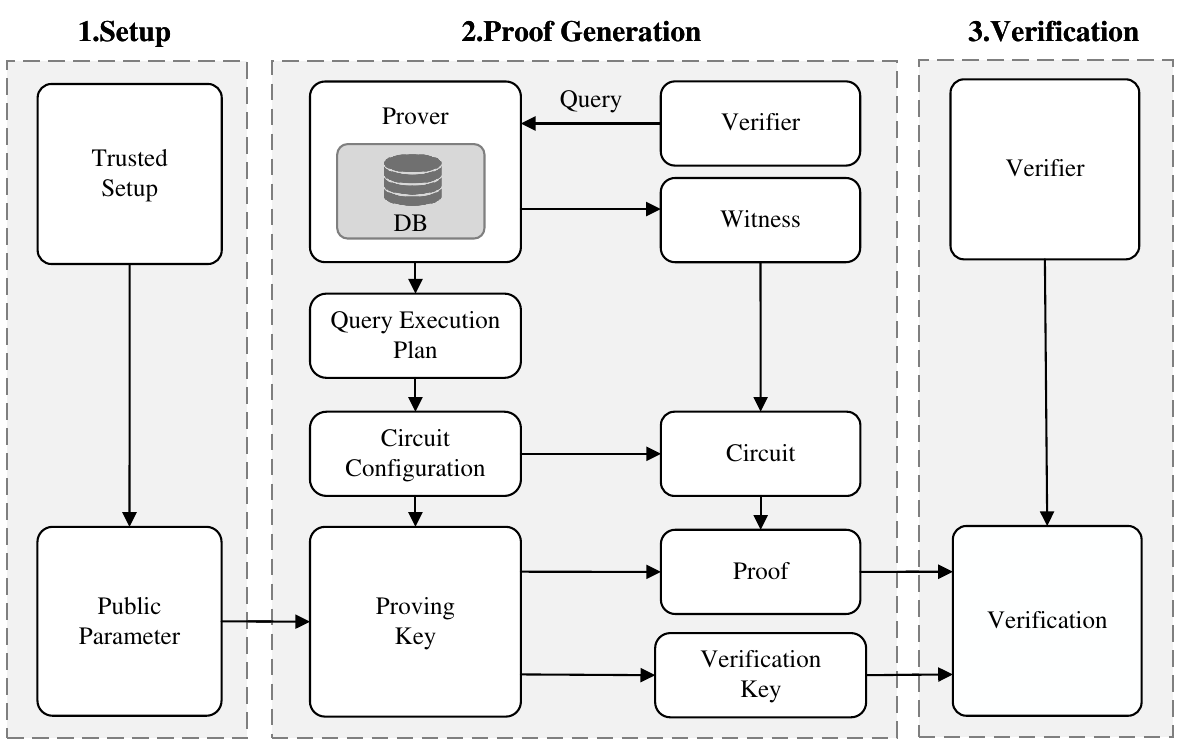}
    \caption{ZKGraph workflow.}
    \label{fig:workflow}
\end{figure}

ZKGraph involves three phases as depicted in \fig~\ref{fig:workflow}.

\textbf{(1) One-time Initialization Phase:}
ZKGraph commences with a one-time trusted setup ceremony.
This ceremony generates global public parameters, the size of which is determined by the maximum number of rows the circuit can support, rather than any specific query structure.
These parameters form the foundation for all subsequent operations within the system.

\textbf{(2) Proof Generation Phase:}
This begins when the verifier submits a query to the prover.
The prover first analyzes the query's structure to derive an execution plan, which in turn defines the specific arithmetic circuit required to represent the query's logic.
Based on this circuit definition and the global public parameters, the system then generates a dedicated proving key (PK) and a corresponding verification key (VK).
Subsequently, the prover executes the query over its private database to obtain the result and construct the witness.
This witness comprises the query result itself, along with all private inputs and intermediate computational steps required to satisfy the circuit's constraints.
Finally, the prover uses the PK, the witness, and the circuit to generate a ZKP.

\textbf{(3) Verification Phase:}
In this phase, the verifier utilizes the VK, the public query result, and the received proof to run the verification algorithm.
A successful verification provides the verifier with the guarantee that the prover executed the query faithfully and that the provided result is authentic.

\subsection{Expansion-Centric Model}

Translating complex graph queries into ZKPs presents a significant challenge.
A naive, direct translation of graph queries can result in prohibitively large and inefficient circuits, rendering them impractical.
To overcome this, ZKGraph introduces an ``expansion-centric'' model.
Instead of proving the entire execution of a graph query, this model deconstructs a complex query into a sequence of modular attestations.
This shifts the verification challenge from providing a single complex proof to validating a chain of well-defined, composable steps.
The cornerstone of this model is the \textit{expansion} operation, which involves navigating from a source node to its adjacent nodes, following the connecting edges.
This process of traversing the graph forms the foundation for more complex queries.
Most other graph operations can be conceptualized as extensions or compositions of this fundamental expansion operation, essentially adding specific attributes or constraints to the traversal.

For instance, the expansion operation itself does not inherently possess the constraint of finding the shortest path. 
As such, a specialized version of expansion, \ie the well-known breadth-first search (BFS), is employed.
The BFS systematically explores the graph level by level, guaranteeing that the first time a target node is reached, it is through the shortest possible path in terms of the number of edges.

Similarly, a simple expansion operation lacks a filtering mechanism.
To introduce this capability, a \textit{filter} operator is integrated into the traversal process.
This allows for the selective inclusion or exclusion of nodes and edges based on their properties, thereby refining the query results.
The need for ordered results gives rise to the \textit{order-by} operator, which sorts the output of a traversal based on specified criteria.

Therefore, many distinct graph operations are, in essence, variations of the expansion operation endowed with particular attributes: the ``shortest'' attribute for path-finding, the ``filtering'' attribute for selective traversal, the ``ordering'' attribute for sorted results, the ``bidirectional'' attribute for traversals that consider edge direction, and so on.
Different and more complex queries are then constructed by composing these attribute-laden expansion operations in various sequences.
Such an expansion-centric model aligns well with the verification approach in zero-knowledge circuits. 
Rather than executing the entire graph algorithm directly within the circuit, our approach uses the circuit to verify that the intermediate and final results, computed by the prover, are consistent with both the graph query's constraints and the private graph database.

\subsection{Edge Storage Formats}
\label{subsec:format}

In ZKP graph databases, the efficient representation of data, particularly the edge storage format, is paramount.
The volume of input data is a critical factor here, directly impacting system performance and the overhead associated with proof generation.
Two primary representations for graph edges include adjacency matrices and the compressed sparse row (CSR) format, among others.
Consequently, the CSR format presents a compelling advantage over adjacency matrices, primarily owing to its superior space efficiency for representing sparse graphs, which are prevalent in many real-world scenarios.

Although CSR offers compact storage, which potentially reduces the amount of data to be committed in ZKP, our work on designing operators for zero-knowledge circuits revealed that this format introduces intractable issues.
Specifically, the Compressed Sparse Row (CSR) format is a data structure that represents a sparse matrix using three arrays: one for the non-zero values, a second to define the extents of rows, and a third for the column indices of these values.
The primary issue with CSR in the context of arithmetization for ZKPs is its reliance on indexed data access.
However, such indexing, which implies dynamic memory access patterns based on runtime values, is often antithetical to the static nature required for efficient arithmetization and polynomial commitment schemes.
Consequently, we adopt an edge list format, despite its potentially less compactness, to avoid these complexities.
We will discuss their differences in detail in Section~\ref{subsec:ss_expansion}, where the operator design of the single-source expansion is used as an example.

% These custom circuits are designed for correctness and efficiency in a zero-knowledge context, and are tailored to handle graph operations. 
% \paragraph{Polynomial Commitments}The prover commits to polynomials that encode the essential components of the circuit.
% \paragraph{Vanishing Argument}To demonstrate that all circuit constraints are satisfied, the prover constructs the vanishing argument to ensure that all circuit relations are constrained to zero.
% \paragraph{Polynomial Evaluation}The prover produces the purported evaluations of the various polynomials at challenge points derived via the Fiat-Shamir heuristic.
% \paragraph{Multipoint opening argument}A multipoint opening argument is constructed to efficiently check that the evaluations are consistent with their respective commitments.
% \paragraph{KZG Opening Proof Generation}Finally, the prover generates an opening proof for the multipoint opening argument polynomial using the KZG (Kate-Zaverucha-Goldberg) commitment scheme.

% ZKGraph involves two entities:
% \begin{itemize}
% \item Prover: Possesses the private graph dataset and is responsible for executing queries and generating zero-knowledge proofs of their correct computation.
% \item Verifier: Issues queries and verifies the accompanying proofs to ensure the integrity and correctness of the results received from the Prover.
% \end{itemize}

\section{Operations Design}
\label{sec:oper}

This section elaborates on the design of several core graph operators in ZKGraph.
Following the expansion-centric model proposed in Section~\ref{sec:system}, we deconstruct graph queries into a series of basic and composable verification primitives.
To achieve this, our methodology utilizes the arithmetic PLONKish circuits in Halo2 to store the input, intermediate, and outcomes of queries in the advice columns.
Building upon this data commitment in advice columns, we then develop special custom gates in Halo2.
Each gate is specifically designed to enforce the rules for these fundamental graph operations on the data held within the columns.

\subsection{Single-Source Expansion}
\label{subsec:ss_expansion}

We first introduce the single-source expansion operation, the cornerstone of graph database query processing.
This operation enables the exploration of a node's neighborhood by traversing its outgoing edges, forming the basis for more complex graph queries.
A naive implementation that exhaustively checks all potential connections would incur prohibitive circuit complexity.
Alternatively, our design efficiently proves an edge using a permutation argument \cite{gabizon2019plonk}.

By analyzing our design under edge-list and CSR formats (as discussed in Section~\ref{subsec:format}), we also concretely compare and illustrate the distinct impacts that both storage approaches have on ZKP circuits.

\para{Formulation.}
The input to a single-source expansion operation, designed to traverse one-hop relationships from a source node, typically includes a source node identifier $id_{s}$ to traverse from and an edge table $D_{e}$.
The edge table consists of node connections and optional relationship properties and can be represented in different storage formats as follows:
\begin{itemize}
    \item \textbf{Edge-List Format:} Three columns for source node identifiers $A$, target node identifiers $B$, and optional corresponding edge properties $Val$.
    \item \textbf{CSR Format:} Three columns for target node identifiers $Col$ (concatenated for all source nodes), row pointers $Row$ (indicating the start of each source node's edges), and optional corresponding edge properties $Val$.
\end{itemize}
The output is a result table $D_{out}$ with two columns, where each entry signifies a one-hop traversal from $id_{s}$ to $id_{t}$, enabled by an edge in $D_{e}$ (where $id_{s}$ and $id_{t}$ correspond to the node identifiers in $A$ and $B$, respectively).

\para{Circuit Design for Edge-List Format.}
Our design leverages the list format of $D_{e}$.
While this might involve iterating over a larger number of edges compared to a CSR-based approach, it avoids the complexities of dynamic data access within the circuits.
The prover is responsible for generating the witness, which includes $id_{s}$, the edge data $D_{e}$, the result table $D_{out}$, and an auxiliary flag column $C_{fl}$.
The circuit then proves the correctness of $D_{out}$ using permutation arguments.

The following two conditions must be enforced for $D_{out}$:

\textbf{(1) Edge Completeness:}
The circuit must ensure that the prover correctly identifies all edges originating from $id_{s}$.
This is achieved by constraining the auxiliary column $C_{fl}$.
For each row $i$ in the edge table $D_{e}$, the prover sets $C_{fl}[i]$ to $1$ if the source node identifier, $D_{e}.A[i]$, matches the start node $id_{s}$ and to $0$ otherwise.

\textbf{(2) Edge Correctness:}
After establishing the correct selection of edges, we must prove that the multiset of output targets in $D_{out}$ is identical to the multiset of targets from the selected edges in $D_{e}$.
A standard permutation argument is employed for this proof.
Specifically, we store $(id_{s}, id_{t})$ and the selected $(A[i], B[i])$ for which $C_{fl}[i] = 1$ in columns $C_{s}, C_{t}, C_{a}, C_{b}$.
To reduce the two-column check to a single-column check, we use a random challenge $\alpha$ to compress each tuple into a single field element.
We build two new columns, $C_1$ and $C_2$, as follows:
\begin{equation}
\begin{aligned}
    C_1[i] & = C_{s}[i] + \alpha \cdot C_{t}[i] \\
    C_2[i] & = C_{a}[i] + \alpha \cdot C_{b}[i].
\end{aligned}
\end{equation}
We then prove that the multiset of values in $C_1$ is a permutation of the multiset of values in $C_2$, \ie there exists a permutation $\sigma$ such that $C_1[i] = C_2[\sigma(i)]$ for all relevant indices $i$.
This is commonly implemented in PLONKish systems using a running product argument.
With a second random challenge $\beta$, we constrain an accumulator column $Z$ as follows:
\begin{equation}
\begin{cases}
    Z_0 = 1 \\
    Z_{i+1} = Z_i \cdot \frac{C_1[i] + \beta}{C_2[i] + \beta}, \,\forall i \in [0, n - 2] \\
    Z_{n} = 1
\end{cases}
\end{equation}
This construction ensures $\prod_{i=1}^{n}(C_1[i] + \beta) = \prod_{i=1}^{n}(C_2[i] + \beta)$, which, except with negligible probability, implies that the multisets $C_1[i]$ and $C_2[i]$ are identical.

\para{Correctness for Edge-List Format.}
We establish the correctness of the edge-list format in a two-step process. 
First, the constraints on the auxiliary flag $C_{fl}$ isolate the complete multiset of edges, $M_{in}$, that start from $id_{s}$.
Second, a permutation argument rigorously enforces the equivalence between the output multiset $M_{out}$ and the selected input multiset $M_{in}$.
This combined enforcement guarantees both edge completeness (no valid edges are omitted) and correctness (no invalid edges are included), ensuring that $D_{out}$ is a correct and complete representation of the one-hop expansion.

\para{Circuit Design for CSR Format.}
We detail the circuit design for the single-source expansion operation when graph edges are stored in CSR format.
As mentioned previously, CSR represents a graph using three columns: $Col$, $Row$, and optional $Val$.
While CSR is storage-efficient, its reliance on indexed data access poses great challenges for ZKP.
The dynamic data access patterns of the CSR conflict with the static nature of arithmetic circuits, necessitating a more complex design.

To construct a valid ZKP, the prover must generate a witness that includes not only the primary inputs and outputs but also several intermediate values that deconstruct the CSR access logic.
Specifically, for each source node $id_{s}$,
\begin{itemize}
    \item $l_{s}$ and $r_{s}$ that correspond to $Row[idx_{s}]$ and $Row[idx_{s}+1]$, respectively, where $idx_{s}$ is the row index w.r.t.~$id_{s}$.
    These indices define the segment $Col[l_{s}, \dots, r_{s} - 1]$ of $Col$ that contains the indices of target nodes for $id_{s}$.
    \item An auxiliary index column $C_{idx}$, which ranges from $0$ to $\mathrm{len}(col) - 1$ and is used to explicitly refer to the indices in $Col$ within the circuit.
\end{itemize}
To enforce the correctness of $D_{out}$, the prover first checks the correspondence between the row index $idx_{s}$ and the input $id_{s}$.
This is typically achieved via a check over a node lookup table (where the $i$-th entry exactly corresponds to the $i$-th entry in $Row$ and stores its node identifier), verifying whether the entry at $idx_{s}$ indeed contains $id_{s}$.
Subsequently, leveraging the inherent one-to-one structural correspondence between the node lookup table and $Row$, a second lookup is performed to confirm the correctness of $l_{s}$ and $r_{s}$ values.

Next, to guarantee that $D_{out}$ contains exclusively the target nodes in $Col[l_s, \dots, r_s - 1]$, the circuit enforces the following two conditions.
First, for every entry in $Col$ at index $k$, the prover checks whether $k$ falls within the established range $[l_s, \dots, r_s - 1]$.
Based on this check, a binary selector is enforced for each element of $Col$, effectively marking it as ``selected'' (if within the range) or ``not selected'' (otherwise).
Second, a permutation argument is used to prove that the multiset of $id_{t}$ values appearing in the output table $D_{out}$ is identical to the multiset of values from $Col$ for which the aforementioned ``selected'' binary selector was active.
This step ensures that $D_{out}$ contains exclusively all target nodes for $id_{s}$ from the valid range $[l_s, \dots, r_s - 1]$.

\para{Correctness for CSR Format.}
We establish the correctness of the single-source expansion operator for CSR format by proving that the output table $D_{out}$ correctly represents the complete set of target nodes connected to the source node $id_{s}$.
This is achieved by
(1) proving the integrity of $idx_{s}$ and the derived $l_s$ and $r_s$ into $Row$ using lookups, ensuring that the prover targets the correct segment of $Col$;
(2) enforcing via a binary selector that only elements within the valid range $[l_s, \dots, r_s - 1]$ in $Col$ are marked as ``selected'';
and (3) proving, through a permutation argument, that the multiset of $id_{t}$ values in $D_{out}$ is identical to the multiset of these ``selected'' elements from $Col$.
Collectively, these constraints guarantee both completeness (no valid edges are omitted) and correctness (no invalid edges are included).

\para{Edge-List vs.~CSR.}
We compare the performance of zero-knowledge circuits for the single-source expansion operator in the edge-list and CSR formats.
We use the interactive short query 1 in the LDBC SNB benchmark \cite{erling2015ldbc} for evaluation:
\begin{verbatim}
    MATCH (n:Person{id:$personId})
        -[:IS_LOCATED_IN]->(p:City)
\end{verbatim}

As demonstrated in Table \ref{tab:csr_list}, the edge-list format substantially outperforms the CSR format across all metrics, \ie key generation time, proving time, verification time, and proof size.
This performance gap arises from the inherent complexities of the CSR format when translated into zero-knowledge circuits: its indexed access patterns necessitate additional layers of range checks and lookup arguments, thereby increasing circuit complexity.
In contrast, the edge-list format can provide a more direct, ZKP-friendly representation that simplifies circuit design and reduces auxiliary witness elements.
Thus, we adopt the edge-list format for representation in the design of other operators.
This design choice prioritizes proving efficiency over storage compactness, achieving a trade-off that is increasingly favorable as graph queries become more complex.

\begin{table}[t]
    \centering
    \caption{Efficiency comparison between the edge-list and CSR formats.}
    \label{tab:csr_list}
    \setlength{\tabcolsep}{3pt}
    \begin{tabular}{c|cccc}
        \hline
        Format    & Key Gen.~(s) & Proving (s) & Verification (ms) & Proof Size (KB) \\
        \hline
        edge-list & 3.66 & 11.42 & 4.26  & 1.47 \\
        CSR       & 7.81 & 40.36 & 10.83 & 6.22 \\
        \hline
    \end{tabular}
\end{table}

\subsection{Set-Based Expansion}

Building upon the fundamental single-source expansion operator, we introduce the set-based expansion operator.
This operator is designed to efficiently handle traversals from a collection of source nodes.
A naive iterative application of the single-source expansion operator for each node in the source set would result in high proving costs, as its complexity scales linearly with the size of the set.
Therefore, our approach employs a batch verification method that amortizes the proof cost across the entire set of source nodes, performing the verification in a single, efficient pass.

\para{Formulation.}
The input to a set-based expansion operation, designed to traverse the one-hop relationships from multiple source nodes simultaneously, includes the set $ID_{s}$ of source node identifiers and an edge table $D_{e}$, which is represented in an edge-list format with three columns for source node identifiers $A$, target node identifiers $B$, and optional edge properties.
The output is a result table $D_{out}$, where each entry represents an edge $(id_{a}, id_{b})$ that constitutes a valid one-hop traversal, which means that the source node $id_{a}$ must be in $ID_{s}$, and the edge itself must exist in $D_{e}$.

\para{Circuit Design.}
The core of our design is a ``greatest-lower-bound'' matching strategy.
This technique allows the circuit to verify the inclusion of all relevant edges by examining a sorted edge table just once.
To facilitate this, the prover must generate a witness with multiple auxiliary columns, whose correctness is then enforced by a set of custom constraints within the PLONKish circuit.

To prove a set-based expansion from a given set of start nodes $ID_{s}$, the prover first generates the following witness:
\begin{itemize}
    \item $ID'_{s}$: An extended and sorted version of $ID_{s}$ augmented with minimum and maximum dummy values (\eg $0$ and $max$ based on the domain of node identifiers).
    It must be sorted, \ie $ID'_{s}[i] < ID'_{s}[i+1]$ for each $i$.
    \item $T1$ \& $T2$: Two lookup columns constructed from $ID'_{s}$.
    For each row $i$, $T1[i] = ID'_{s}[i]$ and $T2[i] = ID'_{s}[i+1]$.
    These columns enable the circuit to efficiently verify whether two values are adjacent in $ID'_{s}$.
    \item $D'_{e}$: A sorted version of $D_{e}$ by the source node identifier.
    Let $A'$ and $B'$ be the source and target node columns in $D'_{e}$, respectively.
    \item $C_{aux}$: An auxiliary column with one entry for each row in $D'_{e}$.
    For each $A'[i]$ in $D'_{e}$, $C_{aux}[i]$ stores the largest value from $ID'_{s}$ such that $C_{aux}[i] \leq A'[i]$.
    \item $C'_{aux}$: An auxiliary column w.r.t.~$C_{aux}$. $C'_{aux}[i]$ stores the value that immediately follows $C_{aux}[i]$ in $ID'_{s}$.
    \item $C_{fl}$: A binary flag column. $C_{fl}[i]$ is $1$ if the corresponding row in $D'_{e}$ is selected and $0$ otherwise.
\end{itemize}

To enforce the correctness of the output table $D_{out}$, the prover first validates the integrity of the auxiliary columns $C_{aux}$ and $C'_{aux}$, which are crucial for the greatest-lower-bound logic.
This is achieved by enforcing a bracketing condition $C_{aux} \leq A' < C'_{aux}$.
Its validity is proven by a combination of lookup arguments and custom gates: a lookup argument into $(T1, T2)$ confirms that $(C_{aux}[i], C'_{aux}[i])$ represents a valid pair of consecutive elements from $ID'_{s}$, while range-check gates confirm the ordering.

Following this, the circuit ensures that $D_{out}$ exclusively contains edges originating from the start set $ID_{s}$.
For every row $i$ in $D'_{e}$, a binary flag $C_{fl}[i]$ acts as a selector to mark an edge as ``selected'' or ``not selected''.
The circuit enforces that an edge can only be selected if its source node $A'[i]$ is equal to its verified greatest-lower-bound $C_{aux}[i]$.
This is achieved by constraining the difference $A'[i] - C_{aux}[i]$ to be $0$ if and only if $C_{fl}[i] = 1$.
Finally, a permutation argument is used to prove that the multiset of edges appearing in the output table $D_{out}$ is identical to the multiset of rows from $D'_{e}$ for which this $C_{fl}$ selector was active.

\para{Correctness.}
The correctness of the set-based expansion operator is established by demonstrating that the output table $D_{out}$ correctly and completely represents the one-hop expansion for all nodes linked to the set of source nodes $ID_{s}$.
For \emph{completeness}, we ensure that all valid edges connected to any node in $ID_{s}$ are included in $D_{out}$.
Our greatest-lower-bound matching strategy guarantees that for any edge whose source node $id_{a}$ is in $ID_{s}$, the bracketing constraints will uniquely determine $C_{aux}$ to be equal to $id_{a}$.
Consequently, the edge selection logic correctly sets $C_{fl}[i]$ to $1$, indicating that the edge has been selected.
For \emph{correctness}, we guarantee that only valid edges connected to the nodes in $ID_{s}$ appear in $D_{out}$.
The binary flag column $C_{fl}$ marks edges as ``selected'' only when the corresponding row in $D'_{e}$ is selected.
And the subsequent permutation argument ensures that the output table $D_{out}$ is a permutation of the rows from $D'_{e}$ where $C_{fl}[i] = 1$.
The correctness of the above two properties ensures the correctness of the set-based expansion operator.

\begin{figure}[t]
    \centering
    \includegraphics[width=0.75\linewidth]{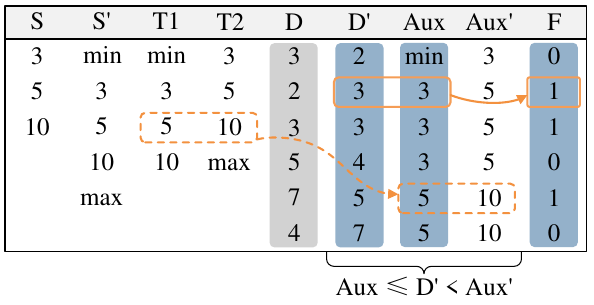}
    \caption{Illustration of the circuit design for set-based expansion.}
    \label{fig:set_based_expand}
\end{figure}

\begin{example}
\fig~\ref{fig:set_based_expand} illustrates the set-based expansion operation, which finds all one-hop neighbors for a given set of source nodes $S = {3, 5, 10}$. 
First, the prover creates an extended and sorted version of the start set, $S'$, by adding $min$ and $max$ as boundary markers.
The source nodes of the edge table are also sorted and placed in column $D'$.
The core of the verification strategy relies on the auxiliary columns $Aux$ and $Aux'$.
For each entry $D'[i]$, the prover computes $Aux[i]$ as the greatest value in $S'$ that is less than or equal to $D'[i]$.
$Aux'[i]$ is then assigned as the immediate successor of $Aux[i]$ from $S'$.
The circuit verifies the integrity of this assignment by checking the bracketing condition $Aux \leq D' < Aux'$.
Finally, the prover generates a binary flag column $F$, where $F[i]$ is set to $1$ if and only if $D'[i]$ is equal to $Aux[i]$, which confirms that the edge's source node is a member of the original start set $S$.
The final result of the expansion is then constructed from the rows where the flag column $F$ is 1.
\end{example}

\subsection{Single-Source Shortest Path}
\label{subsec:sssp}

Next, we present our approach to proving the correctness of the single-source shortest path (SSSP) operation.

\para{Formulation.}
The input of the SSSP operation consists of a source node $id_{s}$, an edge table $D_{e}$, and a column $C_{nid}$ that contains all unique node identifiers, including $id_{s}$.
The output is a column $C_{dist}$ w.r.t.~$C_{nid}$ that contains the shortest path distances from $id_{s}$ to all reachable nodes.
For each node that cannot be reached from $id_{s}$, its distance from $id_{s}$ is set to a pre-defined maximum distance $d_{max}$.

\para{Circuit Design.}
The process of proving the SSSP operation begins by running any shortest-path algorithm, such as breadth-first search (BFS) or Dijkstra's algorithm, on the input graph.
Note that the prover can be generalized to any algorithm as long as its result is exact.
The prover maintains the following intermediate witnesses locally:
\begin{itemize}
    \item $I_{s}$: A binary flag column indicating whether each node in $C_{nid}$ is the source node $id_{s}$.
    \item $C_{dist}$: The computed shortest distance from $id_{s}$ to every node in $C_{nid}$. If a node is unreachable from $id_{s}$, its corresponding distance is set to the predefined $d_{max}$. For $id_{s}$ itself, we always have $C_{dist}[id_{s}] = 0$.
    \item $C_{pre}$: For each reachable node $v$ in $C_{nid}$, $C_{pre}$ stores the identifier of the node that immediately precedes $v$ on the shortest path from $id_{s}$.
    \item $C_{pd}$: For each reachable node $v$ in $C_{nid}$, $C_{pd}$ stores the shortest distance from $id_{s}$ to its predecessor node in $C_{pre}$.
    \item $C_{ud}$: For each edge $(u,v)$ in $D_{e}$, $C_{ud}$ stores the computed shortest distance from $id_{s}$ to its source node $u$.
    \item $C_{vd}$: For each edge $(u,v)$ in $D_{e}$, $C_{vd}$ stores the computed shortest distance from $id_{s}$ to its target node $v$.
\end{itemize}

To guarantee the correctness of the SSSP operation, a set of constraints must be met.
Next, we describe these constraints by categorizing them into node- and edge-level constraints.

At the node level, several constraints are enforced for each node identifier in $C_{nid}$.
First, the flag $I_{s}$ is examined to confirm that it correctly identifies the source node $id_{s}$, and a constraint checks that for $id_{s}$, its distance in $C_{dist}$ is $0$.
Second, for any node $v$ not designated as the source node (\ie $I_{s}[v] = 0$), its computed distance $C_{dist}[v]$ is checked against the following rules: $C_{dist}[v]$ must be $C_{pd}[v] + 1$ or $d_{max}$.
In addition, we prove the correctness of the predecessor node witness (\ie $C_{pre}$ and $C_{pd}$). 
A lookup argument is used to verify that the predecessor node $C_{pre}[v]$ and its associated distance $C_{pd}[v]$ for any node $v$ correspond to an entry in the main node-distance mapping (\ie $(C_{pre}[v], C_{pd}[v])$ must be found within the table formed by $(C_{nid}, C_{dist})$).
To ascertain that the paths implied by the $C_{pre}$ column are structurally valid within the input graph, another lookup argument is employed.
This lookup aims to confirm that for any reachable, non-source node $v$, the edge $(C_{pre}[v], v)$ exists in $D_{e}$.

At the edge level, constraints are applied to each edge $(u,v)$ in $D_{e}$.
The witness columns $C_{ud}$ and $C_{vd}$ associated with each edge are checked for consistency with the column $C_{dist}$ using two lookup arguments:
The pair $(u, C_{ud}[u])$ must be found within $(C_{nid}, C_{dist})$, and similarly $(v, C_{vd}[v])$ must also be found within $(C_{nid}, C_{dist})$.
Critically, the relaxation property of the shortest path is proved: for every edge $(u, v)$, the distance to the target node $v$ must be no greater than the distance to the source node $u$ plus one.

In combination, the above node- and edge-level constraints are designed to prove that the prover's witness values of $C_{dist}$ faithfully represent the distances of the shortest-path solution from $id_{s}$.

\para{Correctness.}
The correctness of the SSSP operation is established by demonstrating that if all the constraints detailed above are satisfied by the prover's witnesses, the computed column $C_{dist}$ will contain the shortest path distances from $id_{s}$.
This is established by proving the following three properties.

\textbf{(1) Correct Source Node Initialization:}
The constraints ensure that the flag column $I_{s}$ is set correctly for the source node $id_{s}$.
The constraint $I_{s} \cdot C_{dist} = 0$ forces $C_{dist}[id_{s}] = 0$.

\textbf{(2) Valid Path Construction and Distance Propagation:}
For any node $v$ not flagged as the source, the constraints ensure that $C_{dist}[v]$ is either $C_{pd}[v] + 1$ or $d_{max}$.
The lookup constraint, which verifies that $(C_{pre}$, $C_{pd})$ is a valid pair from $(C_{nid}, C_{dist})$, confirms that $C_{pd}[v]$ is the actual distance of the claimed predecessor $C_{pre}[v]$.
Furthermore, the lookup constraint, which checks the existence of the edge $(C_{pre}[v], v)$ in $D_{e}$, ensures that any claimed predecessor relationship corresponds to an actual edge in the graph.

\textbf{(3) Relaxation Property:}
At the edge level, for every edge $(u, v)$ in $D_{e}$, the lookups of $(u, C_{ud}[u])$ and $(v, C_{vd}[v])$ in $(C_{nid}, C_{dist})$ ensure that the reported distances are consistent with the shortest distances claimed by the prover in $C_{dist}$ for nodes $u$ and $v$, respectively.
Then, we check the fundamental Bellman-Ford relaxation condition, \ie the constraint $C_{vd}[v] \leq C_{ud}[u] + 1$ must hold for every edge $(u, v)$ in $D_{e}$.
If there were a shorter path to any node $v$ than the one reported in $C_{dist}[v]$, there would exist some edge $(x,v)$ for which $C_{dist}[v] > C_{dist}[x] + 1$, thus violating the constraint.
Thus, the satisfaction of the relaxation constraint across all edges, combined with the path construction properties, confirms that the prover does not miss a shorter path than the reported one.

In summary, the satisfaction of these three properties collectively guarantees that the column $C_{dist}$ accurately reflects the actual shortest path distances from $id_{s}$.

\begin{figure}[t]
    \centering
    \includegraphics[width=0.9\linewidth]{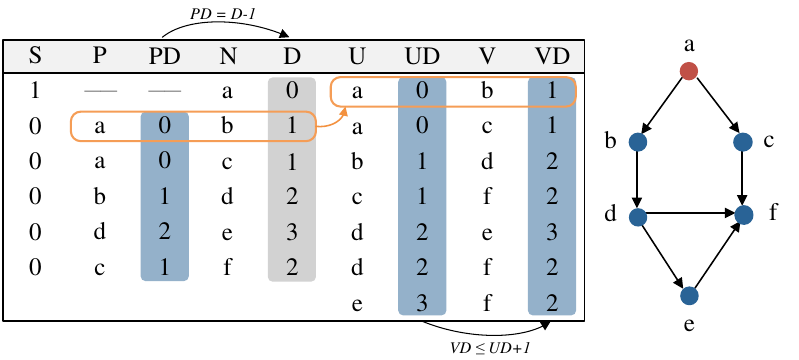}
    \caption{Illustration of the circuit design for single-source shortest path.}
\label{fig:Single-Source Shortest Path}
\end{figure}

\begin{example}
\fig~\ref{fig:Single-Source Shortest Path} illustrates the SSSP operation, which verifies the shortest path distances from a source node $a$.
In the node table, the prover provides the following for each node in $N$: a source flag in column $S$, its shortest distance $D$; and for non-source nodes: its predecessor $P$ and its predecessor's distance $PD$.
For each edge in $(U, V)$, the prover provides their corresponding distances $UD$ and $VD$.
The core of the verification strategy relies on enforcing the two main constraints shown in the figure.
At the node level, the circuit verifies path consistency by checking that a node's distance $D$ is greater than its predecessor's distance $PD$ by $1$ (as shown by $PD = D-1$), and further validates that $(P, N)$ is an actual edge by using a lookup argument to confirm its existence in the edge table $(U, V)$.
At the edge level, it enforces the relaxation property $VD \leq UD + 1$ for every edge, confirming that no shorter path is neglected.
The satisfaction of these constraints across all witness entries constitutes a valid proof that the distances in column $D$ are correct.
\end{example}

\subsection{Bidirectional Relationship Canonicalization}
\label{subsec:bi}

Next, we introduce our approach to proving the correctness of operations to handle bidirectional relationships via canonicalization.
Many relationships in graphs, such as ``friendship'' (\eg $\text{person\_knows\_person}$), are inherently bidirectional.
To manage such relationships consistently and avoid the data redundancy that arises when treating an edge $(u, v)$ as distinct from $(v, u)$, we employ canonicalization to establish a unique, standardized representation for each bidirectional edge.

\para{Formulation.} 
The input for a bidirectional relationship canonicalization operation consists of an edge table, denoted $D_{be}$.
Each row in $D_{be}$ represents a directed edge.
For an edge denoting a relationship between node $id_a$ and node $id_b$, the table might store it as $(id_a, id_b)$ or $(id_b, id_a)$.
The goal is to produce a canonical representation for each such bidirectional edge.
The canonical form of an edge between $id_a$ and $id_b$ is defined as $(id_{l}, id_{h})$, where $id_{l} = \min(id_a, id_b)$ and $id_{h} = \max(id_a, id_b)$.
This ensures that the relationship between $id_a$ and $id_b$ always has the same representation, regardless of the original order of node identifiers in the input data.

\para{Circuit Design.}
The proof of correct canonicalization of bidirectional relationships begins with generating a witness, which includes the original edge node column $C_{ea}$ and $C_{eb}$ from $D_{be}$ and the canonical edge node column $C_{l}$ and $C_{h}$.
To ensure that $(C_{l}, C_{h})$ is the correct canonical representation of $(C_{ea}, C_{eb})$, a set of constraints must be satisfied for each edge.
First, the order constraint, $C_{l} \leq C_{h}$, proves that the identifier claimed to be smaller is indeed less than or equal to the identifier claimed to be larger, ensuring that the canonical representation is always in the correct order.
Second, the sum invariant constraint, $C_{ea} + C_{eb} = C_{l} + C_{h}$, ensures that the sum of the identifiers in the original edge representation is equal to the sum of the identifiers in the canonical representation.
Third, the product invariant constraint, $C_{ea} \cdot C_{eb} = C_{l} \cdot C_{h}$, ensures that the product of the identifiers in the original edge representation is equal to the product of the identifiers in the canonical representation.

\para{Correctness.}
The correctness of the bidirectional relationship canonicalization operation is established by showing that if the three constraints (order, sum invariant, and product invariant) are satisfied by the prover's witnesses, the pair $(C_{l}, C_{h})$ will be the unique canonical form $(\min(C_{ea}, C_{eb}), \max(C_{ea}, C_{eb}))$.
This correctness stems from a fundamental algebraic property concerning the relationship between the roots and coefficients of a polynomial, as described by Vieta's formulas.
Consider the following two quadratic equations:
\begin{equation}
\label{eq:vieta1}
    x^2 - (C_{ea} + C_{eb}) x + (C_{ea} \cdot C_{eb}) = 0
\end{equation}
\begin{equation}
\label{eq:vieta2}
    x^2 - (C_{l} + C_{h}) x + (C_{l} \cdot C_{h}) = 0
\end{equation}
The roots of Eq.~\eqref{eq:vieta1} are $x_1 = C_{ea}$ and $x_2 = C_{eb}$, while the roots of Eq.~\eqref{eq:vieta2} are $x_1 = C_{l}$ and $x_2 = C_{h}$.
The sum invariant constraint $C_{ea} + C_{eb} = C_{l} + C_{h}$ ensures that the coefficients of $x$ are equal in both equations.
The product invariant constraint $C_{ea} \cdot C_{eb} = C_{l} \cdot C_{h}$ further guarantees that their constant terms are equal.
Since both quadratic equations have identical coefficients, they are the same equation.
Therefore, they must have the same set of roots.
This implies that the pairs of identifiers $(C_{ea},C_{eb})$ and $(C_{l},C_{h})$ are equivalent.

Then, the order constraint $C_{l} \leq C_{h}$ ensures that the roots are always in the correct order.
Thus, the satisfaction of the three constraints collectively guarantees that the witnessed pair $(C_{l}, C_{h})$ correctly and uniquely represents the canonical form of the original bidirectional relationship $(C_{ea}, C_{eb})$.

\begin{figure}[t]
    \centering
    \includegraphics[width=0.75\linewidth]{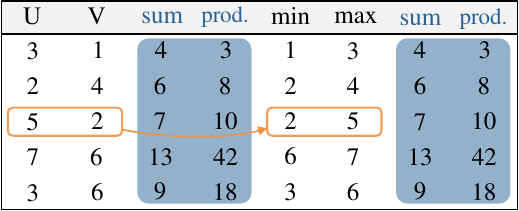}
    \caption{Illustration of the circuit design for bidirectional relationship canonicalization.}
    \label{fig:bi}
\end{figure}

\begin{example}
\fig~\ref{fig:bi} illustrates the process of bidirectional relationship canonicalization, where each input edge $(U, V)$ is transformed into its unique canonical form $(min, max)$.
Verification relies on enforcing three constraints for each row: the order constraint $min \leq max$, and the invariance of both the sum and product $U + V = min + max$ and $U * V = min * max$.
For example, the highlighted edge $(5, 2)$ is correctly transformed to $(2, 5)$ because for both pairs, the sum is 7 and the product is 10, and $2 \leq 5$.
\end{example}

\para{Extension.}
The bidirectional relationship canonicalization operation is not limited to standalone edge processing but also serves as an integrated component for more complex graph query operations, such as expansion and single-source shortest path, when dealing with inherently symmetric relationships.
By creating a unique representation for each bidirectional edge, canonicalization simplifies the circuit design and ensures the correctness of ZKPs for these operations.

\subsection{Other Operations}

In addition to the operations discussed previously, we briefly introduce other operations implemented.

The operations \emph{order-by} and \emph{limit-k} typically appear together and are used to sort the query results and select the top-$k$ entries.
The prover first needs to sort the input data according to the specified ordering column.
Then, auxiliary columns $IS_{k}$ and $IS_{nk}$ are introduced to mark whether each data entry belongs to the selected top-$k$ items.
For example, if selecting the top-20 entries, the prover will identify the value of the 20th item after sorting, which we refer to as $val_{k}$.
The proving process includes the following steps.
First, it uses a lookup argument to ensure that $val_{k}$ originates indeed from one of the entries marked in $IS_{k}$;
Second, for all entries marked in $IS_{k}$, their values must be better than or equal to $val_{k}$ (according to the sorting rule, \eg for descending order, the value is not less than $val_{k}$); and for all entries marked in $IS_{nk}$, their values must be worse than or equal to $val_{k}$.
Furthermore, the constraint $IS_{k} + IS_{nk} = 1$ ensures that each entry is uniquely classified.
Together, these checks ensure the integrity and correctness of the order-by and limit-k operations.

The operation \emph{reachability} is used to determine if a path exists between two nodes in a graph.
If it is claimed that node $s$ can reach node $t$, the prover must provide a sequence of nodes $C_{path} = [n_0, n_1, \dots, n_k]$ as a path witness.
For the path's endpoints, we adopt a more general condition: rather than strictly requiring $s$ to be the start node and $t$ to be the end node of $C_{path}$, we use a lookup argument to ensure that both $s$ and $t$ are present within the provided path witness $C_{path}$.
And for every pair of adjacent nodes $(n_i, n_{i+1})$ in $C_{path}$, a lookup argument checks that this pair constitutes a valid edge in the edge table.
This approach allows the prover to efficiently prove the existence of a path between two nodes, even if the path is not the shortest one.
Thus, the combination of endpoint presence and sequential edge validity confirms the claimed reachability.

The operation \emph{all-shortest-path-enumeration} focuses on identifying all distinct shortest paths between a source node $s$ and a target node $t$.
If the shortest distance from $s$ to $t$ is given as $d$, the problem of enumerating all such shortest paths can be reduced to finding all nodes $p$ such that (1) the shortest distance from $s$ to $p$ is $d-1$ and (2) a direct edge $(p, t)$ exists in the graph.
The prover first utilizes the SSSP operation (as described in Section~\ref{subsec:sssp}) to confirm that a witness column $C_{p}$ correctly and completely represents all nodes at a shortest distance of $d-1$ from $s$.
Then, the prover utilizes a permutation argument to ensure the correct and complete selection of all nodes from $C_{p}$ that possess a direct edge to $t$ in the graph.
This ensures that the prover can efficiently enumerate all shortest paths between $s$ and $t$.

\section{Experiments}

In this section, we first describe the implementation of ZKGraph and the setup for experimental evaluation.
Then, we evaluate the performance of ZKGraph in terms of proving and verification time, memory usage, operator performance, proof size, and scalability.

\begin{table}[t]
    \centering
    \renewcommand{\arraystretch}{1.2}
    \caption{Running times of generating public parameters with varying the maximal number of rows.}
    \label{tab:public_params}
    \begin{tabular}{c|ccccc}
        \hline
        Maximal Number of Rows & $2^{14}$ & $2^{15}$ & $2^{16}$ & $2^{17}$ & $2^{18}$ \\
        \hline
        Running Time (s) & 0.94 & 1.87 & 3.44 & 6.82 & 13.93 \\
        \hline
    \end{tabular}
\end{table}

\begin{table}[t]
    \centering
    \setlength{\tabcolsep}{5pt}
    \caption{Running times of generating proving and verification keys for LDBC SNB queries.}
    \label{tab:key_gen}
    \begin{tabular}{c|ccccccc}
        \hline
        Query & IS3 & IS4 & IS5 & IC1 & IC2 & IC8 & IC13 \\
        \hline
        Running Time (s) & 3.45 & 2.67 & 3.01 & 10.65 & 9.38 & 10.31 & 2.57 \\
        \hline
    \end{tabular}
\end{table}

\subsection{Experimental Setup}

We implemented ZKGraph on top of the Halo2 zero-knowledge proving system \cite{halo2}.
We evaluated ZKGraph on a subset of the LDBC SNB interactive workload \cite{erling2015ldbc}, namely IS3, IS4, IS5, IC1, IC2, IC8, and IC13.
Using these queries, we demonstrate how ZKGraph performs under varying levels of complexity, characterized by the number of operators and data access volumes.
We measure the database scale by the size of the fact table, such as ``person\_knows\_person'' for queries IS3, IC1, and IC13, and ``comment\_hasCreator\_person''  for queries IS4, IS5, IC2, and IC8.
Our results feature three database sizes: 60k, 120k, and 180k, corresponding to the fact tables with 60k, 120k, and 180k rows, respectively.
Using large data instances allows us to probe the scalability of ZKGraph.
By default, our experiments were performed on the data instance with 60k rows.
We deployed ZKGraph on a server running Ubuntu 20.04, equipped with 256GB of memory and two Intel(R) Xeon(R) Silver 4314 CPUs @ 2.40GHz.

\subsection{Results for Setup Cost}

The ZKGraph setup can be divided into two steps.
First, a trusted setup, a preprocessing step that produces a structured reference string, is performed.
This is a one-time computational task that produces parameters that are openly accessible, storable, and broadly reusable for different circuits (provided that they fit within the defined capacity limits).
Table~\ref{tab:public_params} shows the running time for generating these public parameters.

Subsequently, using these established public parameters and defined circuits with witnesses, ZKGraph generates a proving key and a verification key to be used by the prover and the verifier, respectively.
The proving key is used to generate a proof for the given circuit, while the verification key is used to verify the proof.
Table~\ref{tab:key_gen} shows the running time of generating keys for LDBC SNB queries.

\begin{figure}[t]
    \centering
    \subfigure[SSSP vs BFS\label{fig:sssp}]{\includegraphics[width=0.24\textwidth]{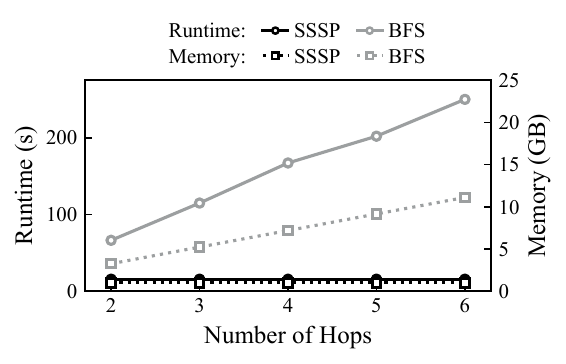}}
    \subfigure[Set expansion\label{fig:expand}]{\includegraphics[width=0.24\textwidth]{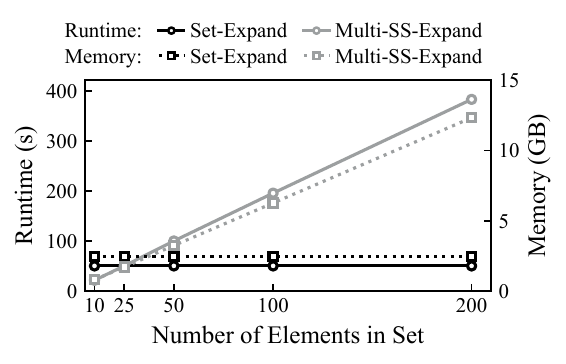}}
    \caption{Performance comparison of different graph algorithms}
    \label{fig:operator_performance}
\end{figure}

\begin{table}[t]
    \centering
    \caption{Effect of bidirectional relationship canonicalization (BiRC) on the performance of set-based expansion and SSSP.}
    \label{tab:Bidirectional}
    \begin{tabular}{c|cc|cc}
        \hline
        \multirow{2}{*}{Method} & \multicolumn{2}{c|}{Set-based Expansion} & \multicolumn{2}{c}{SSSP} \\
        \cline{2-5}
        & BiRC & Preprocess & BiRC & Preprocess \\
        \hline
        Runtime (s) & 8.22 & 21.67 & 26.96 & 31.31 \\
        Memory (GB) & 0.735 & 2.17 & 1.44 & 1.93 \\
        \hline
    \end{tabular}
\end{table}

\subsection{Results for Operators Performance}

To evaluate the performance of operators designed in Section~\ref{sec:oper}, we compare them with the straightforward execution of their corresponding algorithms in ZKP circuits.

First, the single-source shortest path (SSSP) operator is evaluated against a standard breadth-first search (BFS) algorithm implemented directly in circuits.
Since our SSSP operator is designed to verify the computed shortest distances from the source to all other nodes, its performance is independent of the number of hops.
As depicted in \fig~\ref{fig:sssp}, the running times and memory consumption of the SSSP operator remain constant regardless of the increasing hop count.
In contrast, as the number of hops increases from $2$ to $6$, the BFS algorithm requires $1.8 \times$ to $17.4 \times$ longer running time and $3.3 \times$ to $11.3 \times$ higher memory consumption than those of our SSSP operator.
The results highlight the significant performance advantage of our SSSP operator, especially as the traversal depth increases.

\begin{figure}[t]
    \centering
    \includegraphics[width=.8\linewidth]{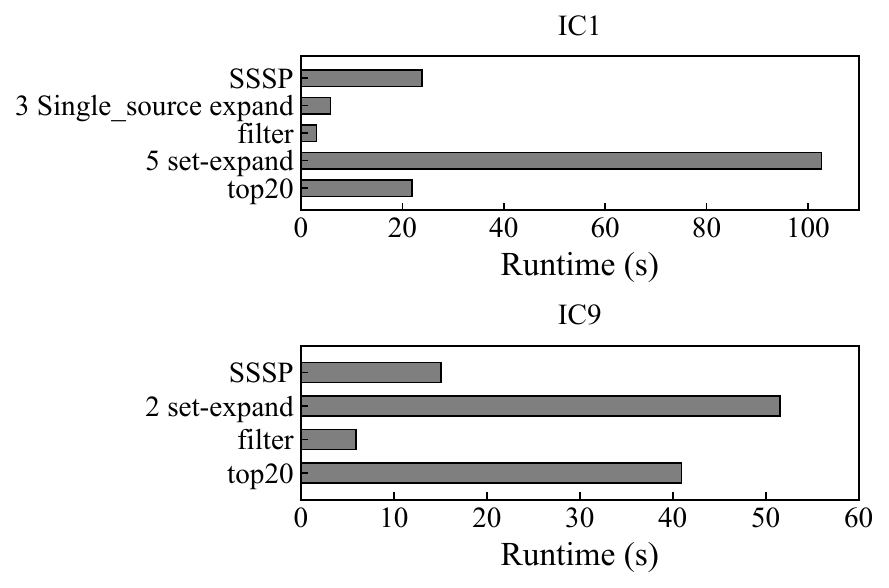}
    \caption{Runtime breakdown of the proof generation steps for IC1 and IC9.}
    \label{fig:ic_cut_performance}
\end{figure}

Next, we evaluated our set-based expansion operator.
Similarly to the SSSP operator, the set-based expansion operator is designed for batch processing.
Thus, its performance is independent of the number of nodes within the input set.
The experimental results in \fig~\ref{fig:expand} confirm this, showing that the set-based expansion operator maintains constant proving time (50 seconds) and memory consumption (2.46 GB) when the number of start nodes increases from 10 to 200.
In contrast, an alternative approach of repeatedly executing the single-source expansion operator exhibits proving time and memory usage that scale nearly linearly with the number of start nodes.
Specifically, the proving time increases from 21 to 383 seconds, and the memory consumption increases from 0.818 to 12.33 GB, respectively, when the number of start nodes grows from 10 to 200.
Although the multiple single-source expansion approach is faster for a small number of start nodes (\eg 10), the set-based expansion operator quickly demonstrates higher scalability as the number of start nodes increases.

\begin{figure*}[t]
    \centering
    \includegraphics[width=\linewidth]{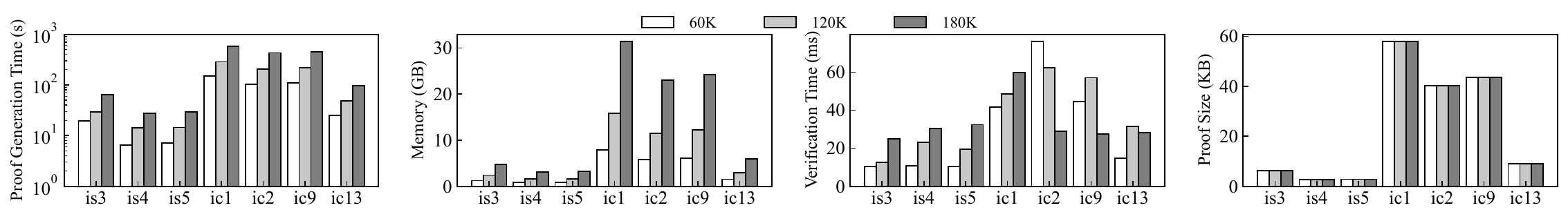}
    \caption{Performance of ZKGraph in terms of proof generation time, verification time, memory usage, and proof size with varying data scales.}
    \label{fig:scalability}
\end{figure*}

We evaluate the performance of our approach when bidirectional relationships are handled by two distinct strategies.
The first strategy is to integrate the canonicalization process described in Section~\ref{subsec:bi} into the circuit design of the set-based expansion and SSSP operators.
The second strategy simply requires the prover to preprocess the witness by expanding each bidirectional edge into two directed edges, $(u, v)$ and $(v, u)$, and then use the standard operator for processing.
As shown in Table~\ref{tab:Bidirectional}, our canonicalization-integrated operators for both set-based expansion and SSSP achieve significant reductions in both runtime and memory usage compared to their preprocessing counterparts.

Moreover, \fig~\ref{fig:ic_cut_performance} illustrates a detailed runtime breakdown of the proof generation process for two representative queries, namely IC1 and IC9. 
We selected these two queries for this analysis because they encompass a comprehensive range of operators designed in this paper, including SSSP, multiple variations of expand operations, filtering, and top-$k$ selection. 
The two figures decompose the total proof generation runtime into the time consumed by each constituent operator, providing insight into performance bottlenecks.

\subsection{Scalability Test}

To evaluate the scalability of ZKGraph, we execute the seven LDBC SNB queries on databases of varying scales, with the fact tables containing 60k, 120k, and 180k rows.
This allows us to analyze ZKGraph's performance when handling larger data volumes.
\fig~\ref{fig:scalability} shows the performance of ZKGraph in terms of proof generation time, verification time, memory usage, and proof size.

On the prover side, both proof generation time and memory usage scale proportionally with the database size.
Such a trend is expected as the sizes of the PLONKish circuits grow linearly with the number of input rows, increasing the prover's computational overhead.
This linear scaling validates our design choice of using low-degree polynomial constraints.

For the verifier, ZKGraph offers exceptional scalability.
\fig~\ref{fig:scalability} demonstrates that both the verification time and the proof size remain nearly constant across all database sizes.
This highly desirable property is a direct result of ZKGraph's adoption of the KZG polynomial commitment scheme, which provides constant-size proofs and constant-time verification.
This makes the verifier's workload minimal and predictable, a critical feature for practical client-side applications.

\section{Related Work}

Verifiable query processing has been studied from multiple perspectives using different techniques.

Trusted execution environment (TEE) based systems \cite{10.1145/3035918.3064030, cryptoeprint:2018/251, 10.1145/3448016.3457308, 201554, 10.14778/2536349.2536353} provide hardware-enforced data confidentiality and integrity guarantees.
However, they are vulnerable to information leakage via memory access patterns, where an adversary can still infer sensitive information by observing memory accesses, even if the data itself is encrypted.
Systems such as Opaque \cite{201554, DBLP:conf/icdt/ArasuK14} address this issue by implementing oblivious operators to hide these patterns, yet at the expense of a larger performance cost.
In contrast, ZKGraph provides verifiability based on cryptographic protocols, avoiding reliance on trusted hardware and its associated trust models.

Methods based on authenticated data structures (ADS) rely on a trusted owner's cryptographic digest, which is built over a structural decomposition of the data, to serve as a trust anchor.
The validity of a query is then ensured by providing a proof composed of specific parts of this pre-computed structure.
These methods typically focus on handling a specific class of computations, such as shortest path \cite{5447914}, pattern matching \cite{6583915,10.14778/2752939.2752944, 10.1145/501983.502003}, range queries \cite{10.1145/1142473.1142488,10.1145/2660267.2660373} and reachability \cite{SONG2023103092,10.1007/3-540-36563-X_20}, when applied to graph data.
In contrast, ZKGraph verifies the correctness of the query evaluation process using composable ZKPs, offering greater flexibility beyond the specific, pre-structured queries typically supported by ADS.

Another line of approaches is based on cryptographic protocols.
For example, vSQL \cite{7958614} and IntegriDB \cite{10.1145/2810103.2813711} use cryptographically verifiable computation to confirm the correctness of SQL queries.
However, these systems were designed for an outsourcing model with the primary goal of ensuring data integrity, which means that they do not inherently provide zero-knowledge privacy guarantees.
While a subsequent extension to vSQL explored the incorporation of ZKP \cite{cryptoeprint:2017/1146}, it does not support ad-hoc queries and fails to address the core challenge of translating arbitrary SQL statements into the cryptographic circuits necessary for such proofs.
Similarly, early work on verifiable graph processing, such as ALITHEIA \cite{10.1145/2660267.2660354}, focuses on providing proofs for specific problems, \eg shortest path and maximum flow, but lacks support for arbitrary graph queries.

Recently, several works have tackled these challenges for relational databases.
ZKSQL \cite{li2023zksql} provides ZKPs for ad-hoc SQL queries using interactive ZKSNARK but still shares the limitations of interactive ZKPs.
To address this, PoneglyphDB \cite{10.1145/3709713} is proposed to evaluate arbitrary SQL queries based on non-interactive ZKPs.
However, the inherent complexity of graph operators introduces new challenges for verifiable computation.
To our knowledge, ZKGraph is the first to generate non-interactive ZKPs for arbitrary graph queries by leveraging our proposed ``expand-centric'' model to deconstruct them into a sequence of modular steps with optimized circuits.

\section{Conclusion}

In this paper, we propose ZKGraph, the first system to leverage non-interactive zero-knowledge proofs for the confidential and verifiable evaluation of graph queries.
To overcome the challenges of prohibitively large and inefficient circuits resulting from a direct translation of graph algorithms, we introduce a novel ``expansion-centric'' graph computation model, which deconstructs complex graph queries into a sequence of modular, composable verification primitives.
Based on this model, ZKGraph implements a suite of highly optimized PLONKish circuits for fundamental graph operations.
Our experimental results demonstrate that ZKGraph significantly outperforms naive approaches, achieving substantial reductions in proving time and memory overhead while ensuring that the verification remains efficient and scalable.

\bibliographystyle{IEEEtran}
\bibliography{refshort}

\end{document}